# Characterizing network paths in and out of the clouds

*Igor Sfiligoi*[1,*], John Graham[1], and Frank Wuerthwein[1]

[1]University of California San Diego, La Jolla, CA 92093, USA

**Abstract.** Commercial Cloud computing is becoming mainstream, with funding agencies moving beyond prototyping and starting to fund production campaigns, too. An important aspect of any scientific computing production campaign is data movement, both incoming and outgoing. And while the performance and cost of VMs is relatively well understood, the network performance and cost is not. This paper provides a characterization of networking in various regions of Amazon Web Services, Microsoft Azure and Google Cloud Platform, both between Cloud resources and major DTNs in the Pacific Research Platform, including OSG data federation caches in the network backbone, and inside the clouds themselves. The paper contains both a qualitative analysis of the results as well as latency and throughput measurements. It also includes an analysis of the costs involved with Cloud-based networking.

## 1 Introduction

Commercial Cloud computing is gaining popularity in the realm of scientific computing. Due to its very flexible nature and large total capacity, it is a great resource for prototyping and also makes for an excellent platform for urgent computing needs. The funding agencies have started to take notice, with several recent grants having explicit mention of commercial Cloud usage [1-4].

An important aspect of any major scientific computing project is data movement. Researchers and their support teams need to understand the basic characteristics of the networks attached to the resources they will use, including latencies and throughput but also cost, in order to make proper planning decisions regarding the resources to use. In the case of commercial Cloud resources, while the performance and cost of compute instances is relatively well documented and understood, the same cannot be said for network links and data movement at large. To address this deficiency, we ran a network characterization campaign in early autumn of 2019, collecting information about throughput and latencies in various regions of Amazon Web Services (AWS), Microsoft Azure and Google Cloud Platform (GCP), both between Cloud resources and major DTNs in the Pacific Research Platform (PRP/TNRP)[5], including Open Science Grid (OSG)[6] data federation caches in the Internet2 network backbone[7], and between different regions inside the Clouds themselves.

---
[*] Corresponding author: isfiligoi@sdsc.edu

Most of the benchmark results are against object storage operated by the three commercial Cloud providers. We use their storage as an endpoint both because it provides an easily accessible scalable endpoint, and because real life data movement often targets those services, too. Section 2 describes performance of data movement inside a single Cloud zone. Section 3 provides information about network performance between Cloud regions, using the Cloud provider's infrastructure. Section 4 describes performance of moving data between Cloud resources and PRP operated hardware. And, finally, section 5 provides an overview of the costs associated with data movement involving commercial Clouds.

## 2 Networking within a Cloud region

In order to measure the performance of the in-region networking, we had to pick a scalable endpoint. Since all commercial Cloud providers operate large pools of distributed storage and offer an easy-to-access object storage interface, we decided that was to be our primary target. Moreover, many scientific compute workloads are likely to access Cloud-native object storage, too, so measuring achievable performance against it was important in its own right.

The test setup was the same for all of the commercial Cloud providers. We created a set of files, each 1 GB in size, and uploaded them to the object storage in one Cloud region per tested commercial Cloud provider, namely one in AWS, one in Azure and one in GCP. We then provisioned a set of compute instances and started a large number of concurrent download processes on them, collecting the timing logs to measure the achieved performance. The tool used to perform the actual download was aria2 [8] and the workload management system was HTCondor [9].

We ran several tests in each of the tested Cloud regions, starting with only a few instances and then progressively increasing the instance count. The scalability of all three tested commercial Cloud providers was excellent, as seen in Figure 1, and we stopped further scaling after we exceeded approximately 1 Tbps aggregate bandwidth.

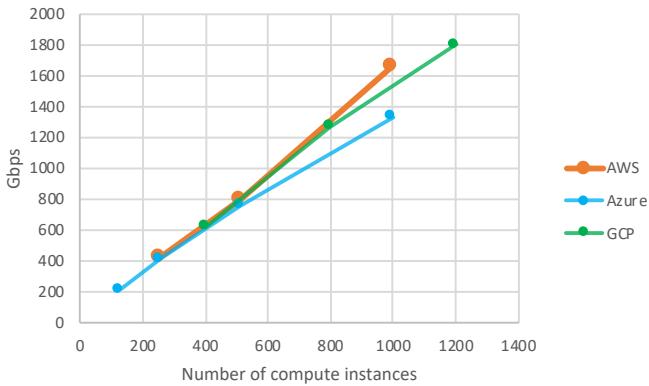

**Fig. 1.** Peak Throughput observed in a Cloud region while downloading from a local object storage instance, for each of the three tested commercial Cloud providers.

## 3 Networking between regions within a Cloud provider

Applications adhering to the distributed High Throughput Computing (dHTC) paradigm, like those currently running on OSG resources, can easily make use of several cloud regions to achieve maximum scalability. Understanding characteristics of network paths between regions is thus very important, too.

The test setup was similar to the in-region one described in Section 2. We tested against Cloud operated cloud storage endpoints but reading from a different Cloud region. We restricted ourselves to tests between regions of the same Cloud provider, since we were mostly interested in network capabilities of the infrastructure operated by those Cloud providers, and not in exercising the peering between them.

The tests showed network throughputs reaching several 100 Gbps on both cross-US and cross-Atlantic links, with GCP touching 1 Tbps. Note that we never exceeded approximately 400 compute instances due to budget constraints, so achievable top speeds on those links for both AWS and Azure are likely significantly higher. The links crossing the Pacific and toward South America did however reach a plateau, showing that those links can barely reach 100 Gbps on AWS and Azure links, but did reach about 1 Tbps on GCP. The observed peak throughputs, alongside ping times, are shown in Tables 1.

**Table 1.** Peak observed throughput and ping times between Cloud regions within a commercial Cloud provider. a) Left, endpoint in the US West region. b) Right, endpoint in the US East region.

|  | Throughput | Ping |  | Throughput | Ping |
|---|---|---|---|---|---|
| AWS East | 440 Gbps | 75 ms | AWS West | 440 Gbps | 75 ms |
| AWS Korea | 100 Gbps | 124 ms | AWS EU | 460 Gbps | 65 ms |
| AWS Australia | 65 Gbps | 138 ms | AWS BR | 90 Gbps | 120 ms |
| AWS Brazil | 80 Gbps | 184 ms |  |  |  |
| Azure East | 190 Gbps | 70 ms | Azure West | 190 Gbps | 70 ms |
| Azure Korea | 110 Gbps | 124 ms | Azure EU | 185 Gbps | 81 ms |
| Azure Australia | 88 Gbps | 177 ms | Azure BR | 100 Gbps | 118 ms |
| GCP East | 1060 Gbps | 68 ms | GCP West | 1060 Gbps | 68 ms |
| GCP Taiwan | 940 Gbps | 119 ms | GCP EU | 980 Gbps | 93 ms |

## 4 Networking between commercial Cloud resources and on-prem

Few science communities spend all of their compute activity inside a Cloud provider. Moving data in and out of the Clouds is thus a normal part of any scientific workflow that uses any amount of commercial Cloud resources. We have thus proceeded in characterizing the networks between the three major Commercial Cloud providers and nodes in the PRP.

Section 4.1 provides information about downloading data from the Cloud-managed object storage into the PRP. Section 4.2 provides information about downloading data from OSG xrootd servers in PRP into Cloud compute instances. Section 4.3 provides a comparison between xrootd and gridftp.

### 4.1 Fetching data from the Clouds

Storing data in Cloud-operated object storage is gaining popularity, due to its ease of use, reliability and scalability. Understanding how fast that data can be retrieved is thus of paramount importance. Note that retrieving data that are the result of computations performed on cloud resources back to the home institutions has similar properties, so the results from this section can be used as a proxy.

Due to the limited number of hardware in the PRP, we designed these tests to measure the performance of network transfers from the Cloud-managed object storage to a single node in the PRP at a time. Even though a single client node was used, multiple concurrent download processes were used. And like the tests in Sections 2 and 3, the tools used were aria2 and HTCondor. Note that all nodes used in the PRP had a 100 Gbps network interface.

The test results were generally quite good, with most tests peaking in the 20 Gbps to 30 Gbps range, as shown in Table 2. Some links were however initially significantly slower, typically in the 1 Gbps to 2 Gbps range. After closer examination, all such links were shown to have improper routing in place, and a simple configuration change was all that was needed to achieve the expected higher throughput described above. Regular testing has thus shown to be an important part of keeping a healthy networking ecosystem.

Table 2. Observed peak throughput while downloading data from Cloud-managed object storage into a single node in PRP

|  | AWS West | Azure West | GCP West | AWS East | Azure East | GCP East |
|---|---|---|---|---|---|---|
| **PRP West (California)** | 35 Gbps | 27 Gbps | 29 Gbps | 23 Gbps | 21 Gbps | 26 Gbps |
| **PRP Central (Illinois)** | 33 Gbps | 27 Gbps | 35 Gbps | 35 Gbps | 36 Gbps | 36 Gbps |
| **PRP East (Virginia)** | 36 Gbps | 29 Gbps |  | 36 Gbps | 31 Gbps |  |

## 4.2 Fetching data into the Clouds

Most scientific applications require a non-trivial amount of input data to compute the results. While data can be pre-staged to a storage medium close to the compute resources, real-time streaming or just-in-time fetching of the data is often the preferred method, for flexibility and reduced complexity reasons. To measure the performance of that operation mode for Cloud-based compute, we measured the performance of Cloud compute instances fetching data from the xrootd servers, operated by OSG [7], that reside on PRP hardware. Due to budgetary constraints, we limited these tests to AWS and Azure platforms.

The test setup was very similar to that in Section 3, since xrootd can be accessed using the HTTP protocol, and thus looks very similar to object storage from a client point of view. Multiple compute instances on Cloud resources were concurrently downloading files using aria2. As shown in Table 3, the observed throughput was around the 20 Gbps mark for the US Central and US East servers. These are quite similar to those in Section 4.1, which was indeed to be expected given that we were exercising the same network links. The throughput from the US West server node was instead mostly higher, with peaks of over 60 Gbps, but we did not investigate the root cause of the discrepancy. All in all, we believe the test results validate the viability of large-scale data transfers in both directions, and also validate the scalability of the OSG infrastructure.

Table 3. Observed peak throughput while downloading data from OSG-operated xrootd services on PRP hardware into Cloud compute instances. Ping RTT from compute instances to the PRP hardware provided underneath.

|  | AWS West | Azure West | AWS East | Azure East |
|---|---|---|---|---|
| **PRP West (California)** | 37 Gbps<br>34 ms | 14 Gbps<br>32 ms | 63 Gbps<br>63 ms | 66 Gbps<br>58 ms |
| **PRP Central (Illinois)** | 21 Gbps<br>71 ms | 18 Gbps<br>45 ms | 21 Gbps<br>19 ms | 20 Gbps<br>67 ms |
| **PRP East (Virginia)** | 21 Gbps<br>75 ms | 20 Gbps<br>69 ms | 23 Gbps<br>24 ms | 23 Gbps<br>69 ms |

We also extended the tests to characterize the performance of cross-Pacific and cross-Atlantic links, as shown in Table 4. Here we observe significant difference in throughput between the two Cloud providers. The networking to Azure resources from the PRP node in California was remarkably good for such high-latency connections, sustaining data transfers in the 20 Gbps to 40 Gbps range. The transfers to AWS resources from the same mode were instead significantly slower, never exceeding 8 Gbps on the cross-Pacific links. The situation was reversed on the cross-Atlantic link while reading from the server node in Europe, where Azure network significantly underperformed compared to AWS.

Table 4. Observed peak throughput while downloading data from OSG-operated xrootd services on PRP hardware into Cloud compute instances.
Ping RTT from compute instances to the PRP hardware provided underneath.

a) Left, server in US West (California).

| Client region | AWS | Azure |
|---|---|---|
| **Korea** | 7.6 Gbps | 32 Gbps |
| | 135 ms | 135 ms |
| **Australia** | 6.3 Gbps | 44 Gbps |
| | 159 ms | 151 ms |
| **Europe** | 14 Gbps | 24 Gbps |
| | 152 ms | 138 ms |

b) Right, server in Europe (Netherlands).

| Client region | AWS | Azure |
|---|---|---|
| **US East** | 12 Gbps | 6.3 Gbps |
| | 94 ms | 81 ms |
| **Europe** | 11 Gbps | 13 Gbps |
| | 7 ms | 3 ms |

### 4.3 Comparing xrootd over HTTP against GridFTP

GridFTP has long been the workhorse of scientific data movement. But access to most of the Cloud-operated storage is based on HTTP-based tools, as already mentioned in the previous sections. In order to validate the switch to the alternate protocol, we measured the achievable throughput of a set of client processes on a single Cloud instance against both a GridFTP server and a xrootd server over HTTP. Both server processes were running on the same hardware.

When reading from a single on-prem server into a single Cloud instance, neither the xrootd nor GridFTP clearly dominate, as can be seen in Table 5, with GridFTP winning some and xrootd over HTTP winning others. Since the tests were run at different times, on shared Cloud resources and over shared network links, some variation is to be expected. We thus consider the two methods of fetching data to be essentially equivalent in terms of performance.

Table 5. Observed throughput downloading a 10GB file from a single on-prem server into a single Cloud instance.

| | GridFTP | xrootd over HTTP |
|---|---|---|
| **GCP US West from UCSD** | 8.9 Gbps | 7.2 Gbps |
| **AWS US West from UCSD** | 9.3 Gbps | 8.4 Gbps |
| **AWS US East from PRP US NY** | 11 Gbps | 21 Gbps |
| **AWS EU Central from PRP EU NL** | 13 Gbps | 11 Gbps |

## 5 Commercial Cloud networking cost

Unlike most on-prem network infrastructures, networking is a billable entity in the commercial Clouds, with the final user being responsible for any and all the costs associated with data movement. It is thus important to understand both the cost model itself and who is billed when the two endpoints are associated with different users.

Fortunately, all major commercial Cloud providers have a very similar cost model, with only minor differences in cost between them. At a high level, network traffic can be split in four distinct categories:
1) Network traffic which stays in the same Cloud zone.
2) Incoming network traffic, from either another Cloud zone of the same Cloud provider or from anywhere else, including the public internet.
3) Network traffic leaving a Cloud zone for another zone of the same Cloud provider.
4) Network traffic leaving the Clouds toward any other destination, including the public internet.

The first two types of traffic, i.e. all incoming traffic, is always free. Importing data from on-prem into the Clouds does not incur any Cloud costs. For the purpose of these tests we have imported over 30 TB of data and were not charged anything for it.

The last two types of traffic are instead both billable, although at a different rate. While the exact amount varies slightly between Cloud providers and affected geographical regions, the traffic between zones inside the same Cloud provider is typically priced at between $10 and $20 per TB, and traffic leaving a Cloud provider's network is typically billed at $50-$90 per TB. Note that this implies that networking related to any and all data exported from Cloud storage to either a different region or to the outside world is billed to the data owner, not the data consumer.

However, researchers do not need to pay for all billable networking. The three major Cloud providers have agreements in place with most research and academic institutions that waive the networking costs of up to 15% of the total monthly bill. This means that even outgoing networking is free for most compute-heavy workloads. To put in perspective, one GB of network leaving the Cloud provider is waved approximately every 15 CPU hours of compute, or every 30 minutes of usage of one V100 GPU.

## 6 Conclusions

This paper provides a snapshot in time of what networking of the major Cloud providers, namely AWS, Azure and GCP, is capable of and shows that such networking is capable of sustaining high throughputs both inside a single Cloud zone and between zones residing on different continents. Network speeds in excess of 100Gbps are commonplace, with several network tests exceeding 1 Tbps aggregate throughputs.

A snapshot of network throughput between the Cloud providers and on-prem nodes operated as part of the Pacific Research Network is also provided. While such throughput is generally not as large as the one observed inside the Cloud networks themselves, it still exceeded 20 Gbps on most network tests. That said, several misconfigured routes have been found during initial testing, and while they were easy to fix, it did show that regular network testing toward cloud resources is necessary in this age of Cloud computing.

Cost of data movement for data leaving Cloud resources is definitely something every Cloud user should be aware of, but such costs are typically not a problem for any compute intensive workflow. And incoming traffic is free.

This work has been partially funded by NSF grants OAC-1826967, OAC-1541349, OAC-1841530, OAC-1836650, MPS-1148698 and OAC-1941481. A fraction of the Cloud costs were covered by Cloud credits provided by Amazon and Microsoft.